\documentclass[]{aa}

\usepackage{graphicx}
\usepackage{natbib}
\bibpunct{(}{)}{;}{a}{}{,}

\title{
The {\sl ISO} 170~$\mu$m luminosity function of galaxies
}

\author{
  T.~T.~Takeuchi\inst{1}\thanks{
  Postdoctoral Fellow of the Japan Society for the Promotion of Science (JSPS)
  for Research Abroad.}
  \and
  T.~T.~Ishii\inst{2}\thanks{Postdoctoral Fellow of the JSPS.}
  \and
  H.~Dole\inst{3}
  \and
  M.~Dennefeld\inst{4}
  \and
  G.~Lagache\inst{3}
  \and
  J.-L.~Puget\inst{3}
}

\institute{
  Laboratoire d'Astrophysique de Marseille, 
  Traverse du Siphon BP8, F-13376 Marseille Cedex 12, France\\
  \email{tsutomu.takeuchi@oamp.fr}
  \and
  Kwasan Observatory, Kyoto University, Yamashina-ku, Kyoto
  607--8471, Japan
  \and
  Institut d'Astrophysique Spatiale, b\^{a}t 121, Universit\'{e} Paris-Sud,
  F-91405 Orsay Cedex, France
  \and
  Institut d'Astrophysique de Paris, 98bis Bd Arago, F-75014 Paris, France
}

\offprints{T. T. Takeuchi\\
\email{tsutomu.takeuchi@oamp.fr}.
}

\date{Received/Accepted}

\titlerunning{170~$\mu$m galaxy luminosity function}
\authorrunning{T.\ T.\ Takeuchi et al.}

\begin{document}

\abstract{
We constructed a local luminosity function (LF) of galaxies using a 
flux-limited sample ($S_{170} \ge 0.195~\mbox{Jy}$) of 55~galaxies 
at $z < 0.3$ taken from the {\sl ISO} FIRBACK survey at 170~$\mu$m.
The overall shape of the 170-$\mu$m LF is found to be different from that of 
the total 60-$\mu$m LF \citep{takeuchi03b}:
the bright end of the LF declines more steeply than that of 
the 60-$\mu$m LF.
This behavior is quantitatively similar to the LF of the cool subsample
of the {\sl IRAS} PSC$z$ galaxies.
We also estimated the strength of the evolution of the LF by assuming the 
pure luminosity evolution (PLE): $L(z) \propto (1+z)^Q$.
We obtained $Q=5.0^{+2.5}_{-0.5}$ which is similar to the value obtained
by recent {\sl Spitzer} observations, in spite of the limited sample size.
Then, integrating over the 170-$\mu$m LF, we obtained the local luminosity 
density at $170\;\mu$m, $\rho_L(170\mu\mbox{m})$.
A direct integration of the LF gives 
$\rho_L(170\mu\mbox{m}) = 1.1 \times 10^8 h \;L_\odot \mbox{Mpc}^{-3}$, 
whilst if we assume a strong PLE with $Q=5$, 
the value is $5.2 \times 10^{7} h \; L_\odot \mbox{Mpc}^{-3}$.
This is a considerable contribution to the local FIR luminosity density.
By summing up with other available infrared data, we obtained the total 
dust luminosity density in the Local Universe, 
$\rho_L(\mbox{dust})=1.1 \times 10^{8}h\;L_\odot \mbox{Mpc}^{-3}$.
Using this value, we estimated the cosmic star formation rate (SFR) density 
hidden by dust in the Local Universe.
We obtained $\rho_{\rm SFR}(\mbox{dust}) \simeq
1.1\mbox{--}1.2\, h \times 10^{-2} \; M_\odot \,\mbox{yr}^{-1}
\mbox{Mpc}^{-3}$, which means that 58.5~\% of the star formation is 
obscured by dust in the Local Universe.
\begin{keywords}
  dust, extinction --- galaxies: evolution --- galaxy formation --- 
  galaxies: luminosity function, mass function --- infrared: galaxies
\end{keywords}
}

\maketitle

\section{Introduction}

The luminosity function (LF) of galaxies is one of the fundamental 
statistics to describe the galaxy population in the universe.
The far-infrared (FIR) LF is vitally important to evaluate the
amount of energy released via dust emission, and further, the fraction of
the star formation activity hidden by dust 
\citep[e.g.,][]{perez_gonzalez05,lefloch05,takeuchi05c}.
Not only the local LF but also its evolution plays a crucial role in 
understanding the cosmic star formation history.

Most of the previous LF works at mid-infrared (MIR) and FIR wavelengths
have been made based on {\sl IRAS} database 
\citep[see,][among others]{rieke86,lawrence86,rowan_robinson87,soifer87,
saunders90,isobe92,rush93,koranyi97,fang98,shupe98,springel98,takeuchi03b}.
The longest wavelength band of the {\sl IRAS} is $100\,\mu$m.
Subsequently, 15 and $90\,\mu$m LFs have been presented based on 
{\sl ISO} data \citep{xu00,serjeant01,serjeant04}.
Now, by the advent of {\sl Spitzer},\footnote{URL: 
{\tt http://www.spitzer.caltech.edu/}.} 
MIR(12 or $15\;\mu$m) LFs based on the 24-$\mu$m band started to be 
available up to $z\sim 1$ \citep{perez_gonzalez05,lefloch05}.
Also recently, {\sl Spitzer}-based 60-$\mu$m LF has been presented
\citep{frayer05}.
At $850\;\mu$m, a LF of {\sl IRAS}-selected sample of submillimeter galaxies
has been published \citep{dunne00}.

At wavelengths between $100\,\mu$m and $850\;\mu$m, however, only a very 
limited number of LFs have been studied.
Further, most of them are made from the sample selected at
shorter wavelengths \citep[e.g.,][]{franceschini98,dunne00},
or estimated/extrapolated from LFs of shorter wavelengths, e.g., $60\,\mu$m 
\citep[e.g.,][]{serjeant05}.
A direct construction of the LF is still rarely done up to now 
\citep[see, ][]{oyabu05}.
Hence, it remains an important task to estimate the LF at wavelengths longer
than $100\,\mu$m from a well-controlled deep survey sample.
Wavelengths between $100\;\mu$m and $850\;\mu$m are also very important in
the context of the extragalactic background radiation, particularly
the cosmic infrared background (CIB).
The CIB is now understood as an accumulation of radiation from dust in
galaxies at various redshifts ($z$).
At the FIR, although the measured CIB is very strong 
\citep[e.g., ][among others]{gispert00,hauser01,lagache05}, 
the properties of the sources contributing to the background is rather
poorly known compared with other wavelengths.
Thus, it is also of vital importance to have a LF at FIR for cosmological 
studies.

In this work, we estimate the LF of the local galaxies ($z < 0.3$) 
at $170\;\mu$m based on the data obtained by FIRBACK survey
\citep{puget99}.
This paper is organized as follows: In Section~\ref{sec:data}, we describe
the $170\;\mu$m galaxy sample.
We present the statistical estimation method of the LF in 
Section~\ref{sec:analysis}.
In Section~\ref{sec:results}, we show the LF and discuss its uncertainties.
Section~\ref{sec:conclusion} is devoted to our conclusions.
We provide numerical tables of our LFs in Appendix~A.
Throughout this manuscript, 
we adopt a flat lambda-dominated cosmology with 
$h\equiv H_0/100 [\mbox{km\,s}^{-1}\mbox{Mpc}^{-1}]$, and
$(\Omega_0,\lambda_0)=(0.3,0.7)$, where $\Omega_0$ is the density parameter
and $\lambda_0$ is the normalized cosmological constant.
We denote the flux density at frequency $\nu$ by $S_\nu$, 
but for simplicity we use a symbol $S_{170}$ to represent $S_\nu$ 
at a frequency ($1.76 \times 10^{12}$~Hz) corresponding to $\lambda=170\;\mu$m.

\section{Data}\label{sec:data}

\begin{figure}
\centering\includegraphics[width=0.45\textwidth]{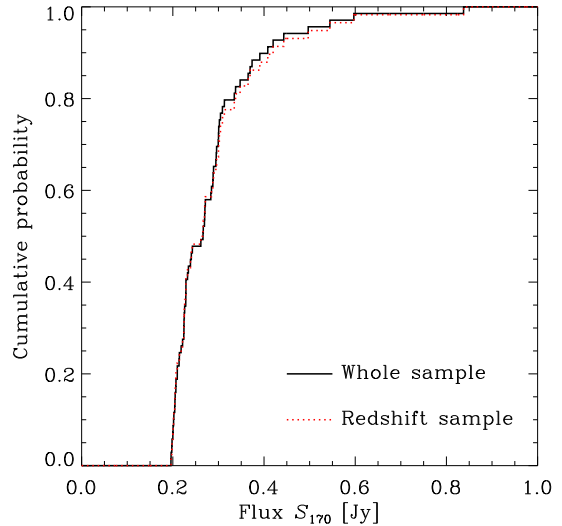}
\caption{The Kolmogorov--Smirnov test for the whole photometric sample
and redshift subsample taken from FIRBACK $170\,\mu$m survey.
The solid line shows the cumulative probability distribution of flux 
densities for the whole sample, while the dotted line depicts that of
the redshift sample.
The difference is found to be very small.
}\label{fig:ks_test}
\end{figure}

\subsection{Parent sample}

The FIRBACK (Far-InfraRed BACKground) survey \citep{puget99,lagache01,dole01} 
is one of the deepest surveys performed at $170\;\mu$m by {\sl ISO} using
ISOPHOT \citep{lemke96}.
It covers $4\;\mbox{deg}^2$ on three fields.
In this work, we use two of these fields: 
FIRBACK South Marano field ($0.89\;\mbox{deg}^2$) and
FIRBACK ELAIS N1 field ($1.87\;\mbox{deg}^2$).\footnote{
The choice of the two fields is due to the follow-up
allotment was different for ELAIS N1/South Marano 
\citep{dennefeld05} and for ELAIS N2 \citep{taylor05}
in the FIRBACK project.
Consequently the conditions of the data aquisition are different between them.
A coherent treatment remains as a future work.
}

The parent sample of FIRBACK is composed of the flux-limited sample of 
141 sources with $S_{170} \ge 135\;\mbox{mJy}$
($3\sigma$ limit).
Flux completeness of this parent sample is $75\;\%$, and 
at flux density $S_{170} \sim 200$~mJy, it becomes $\sim 90$~\% \citep{dole01}.
We use the sources brighter than a flux density of 195~mJy
in the following analysis.

\subsection{Redshifts and completeness}

\begin{figure}
\centering\includegraphics[width=0.45\textwidth]{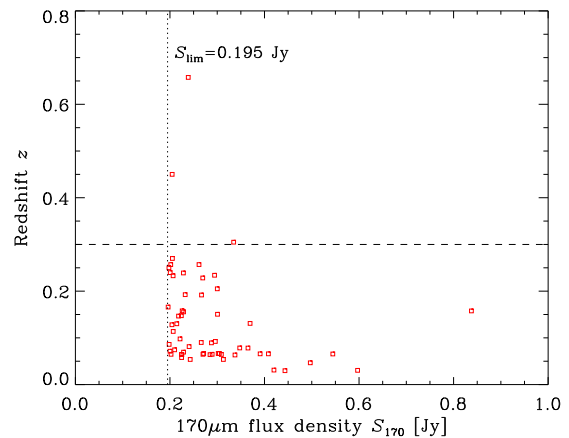}
\caption{The flux density--redshift distribution of our flux-limited
sample.
We used a subsample at $z < 0.3$ to construct the local luminosity 
function (LF).
The vertical dotted line shows $S_{170}=0.195\;\mbox{Jy}$, and 
the horizontal dashed line depicts $z = 0.3$.
}\label{fig:flux_z_dist}
\end{figure}

Redshifts are measured for 58 galaxies out of 69 galaxies above the flux 
density of 195~mJy, i.e., the redshift completeness of 
the sample used is 84~\%.
The redshift measurements have been performed by \citet{chapman02},
\citet{patris03}, and \citet{dennefeld05}.
Since the redshift measurement becomes more difficult toward the fainter
sources, the completeness depends systematically on the flux levels.
Thus, we should examine whether the redshift selection distort
the flux distribution of the sample.

We performed the Kolmogorov--Smirnov test \citep[e.g.,][]{hoel71,hajek99} to
compare the flux-limited sample ($S_{170} \ge 195~\mbox{mJy}$) with the
redshift sample (see Figure~\ref{fig:ks_test}).
The maximum difference between the cumulative distribution functions of
flux-limited and redshift samples are 0.0377.
This shows that we cannot reject the null hypothesis that the two samples 
are taken from the same parent distribution.
Hence, we use the redshift sample as an unbiased subsample of the whole 
flux-limited sample and simply multiply the inverse of the completeness to
obtain the final galaxy density.
The distribution of the flux densities and redshifts of the sample
is shown in Figure~\ref{fig:flux_z_dist}.

The redshift completeness of the sample is also tested by $V/V_{\rm max}$ 
statistics \citep{schmidt68,rowan_robinson68}.
Here $V$ is the volume enclosed in a sphere whose radius is the distance of 
a considered source, and $V_{\rm max}$ is the volume enclosed in a sphere
whose radius is the largest distance at which the source can be detected.
If the sample is complete, $V/V_{\rm max}$ values of the sample galaxies 
is expected to distribute uniformely between 0 and 1, with an average 
$\langle V/V_{\rm max} \rangle =0.5$ and a standard deviation 
$(12n)^{-1/2}$ ($n$: sample size).
For our redshift sample, the mean and standard deviation of the 
$V/V_{\rm max}$ is $0.66 \pm 0.23$, i.e., the sample can be regarded as
complete.
Moreover, it is larger than 0.5 (but within the uncertainty), suggesting the 
existence of evolution
\citep[see, e.g.,][p.444]{peacock99}.

For the estimation of the local LF, we use a subsample of galaxies 
with $z < 0.3$.
The size of this `low-$z$' subsample is 55.
The mean and median redshift of this low-$z$ sample is 0.12 and 0.09, 
respectively.

\section{Analysis}\label{sec:analysis}

\subsection{$K$-correction}

\begin{figure*}
\centering\includegraphics[width=0.45\textwidth]{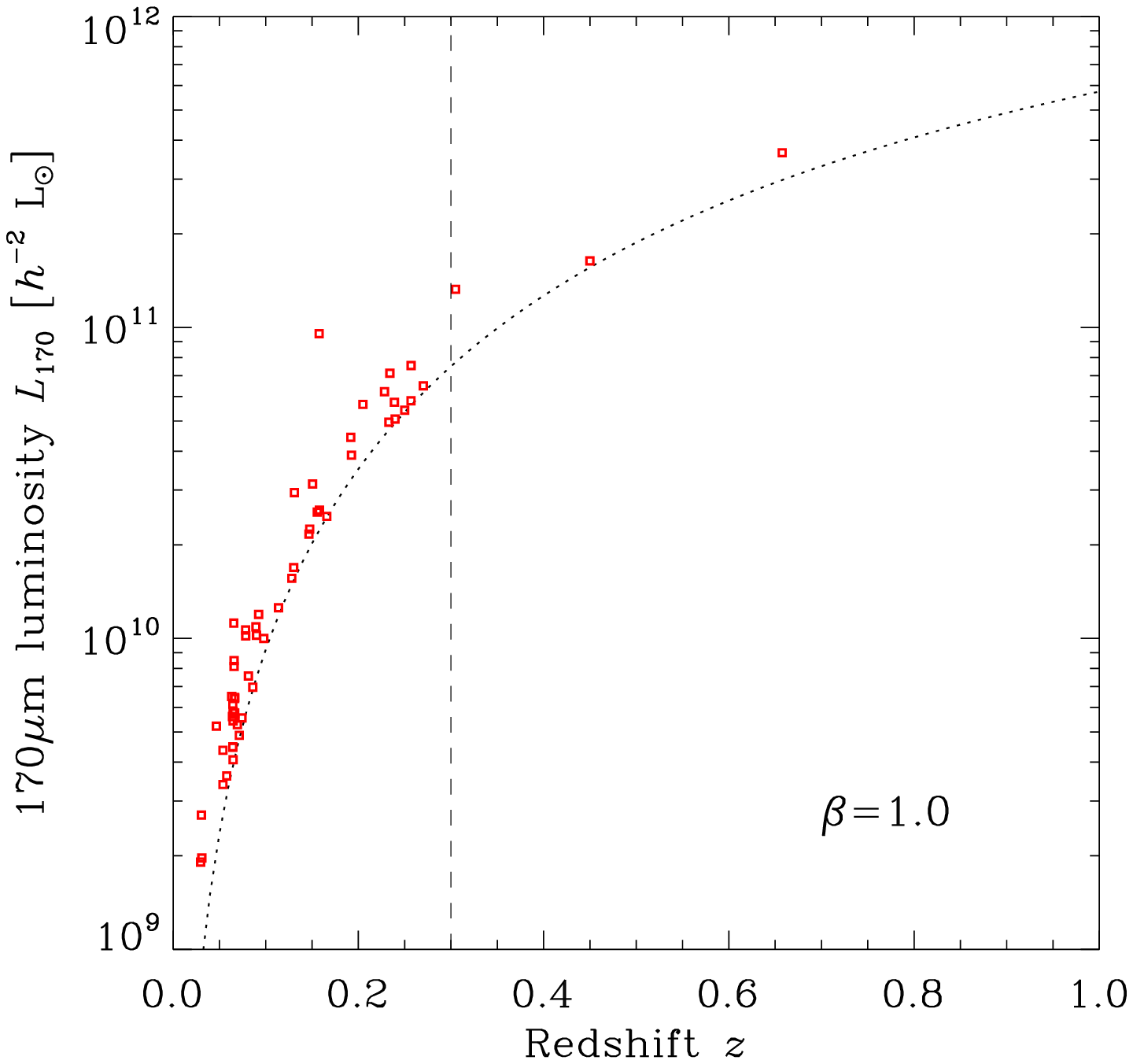}
\centering\includegraphics[width=0.45\textwidth]{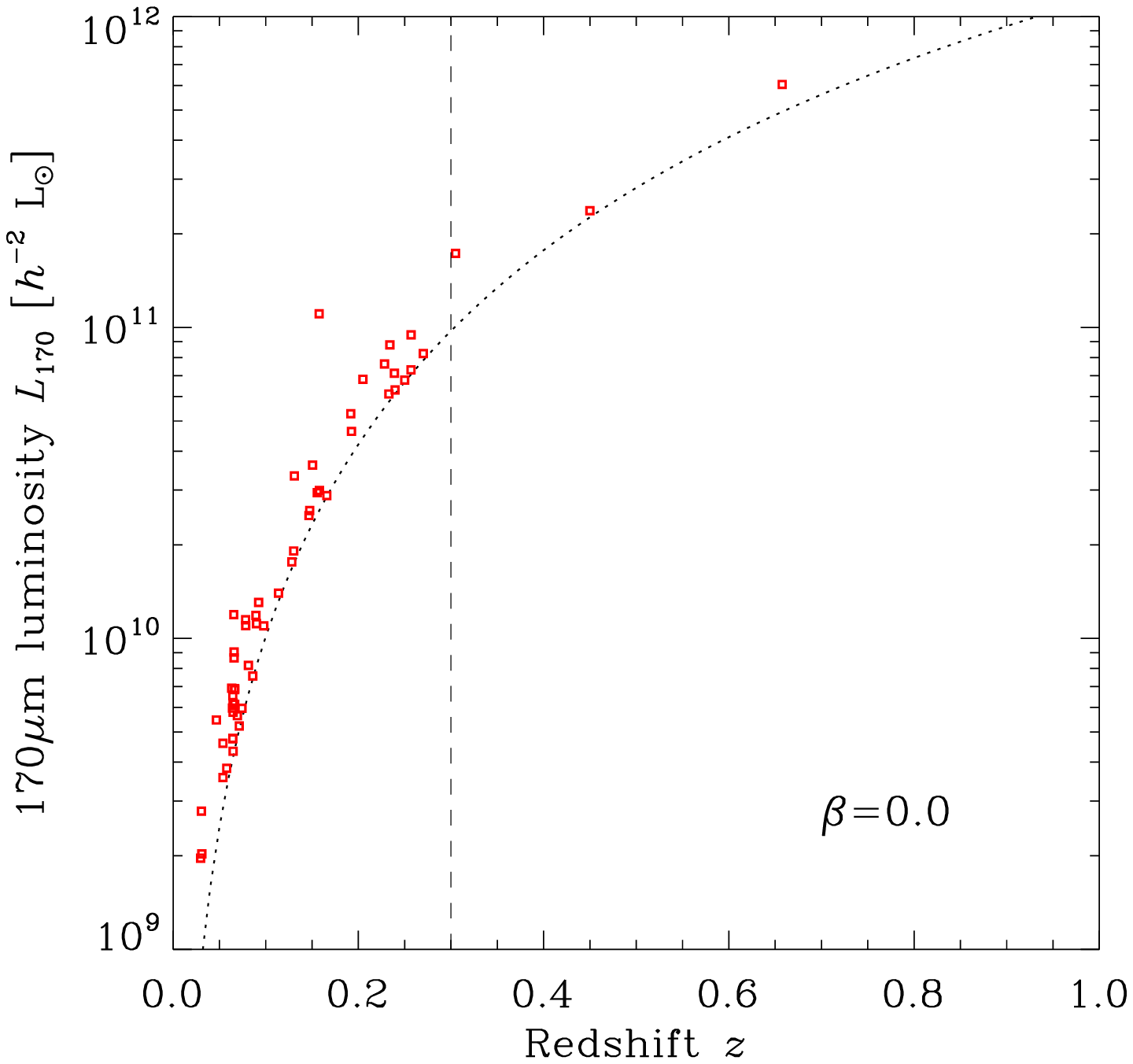}
\centering\includegraphics[width=0.45\textwidth]{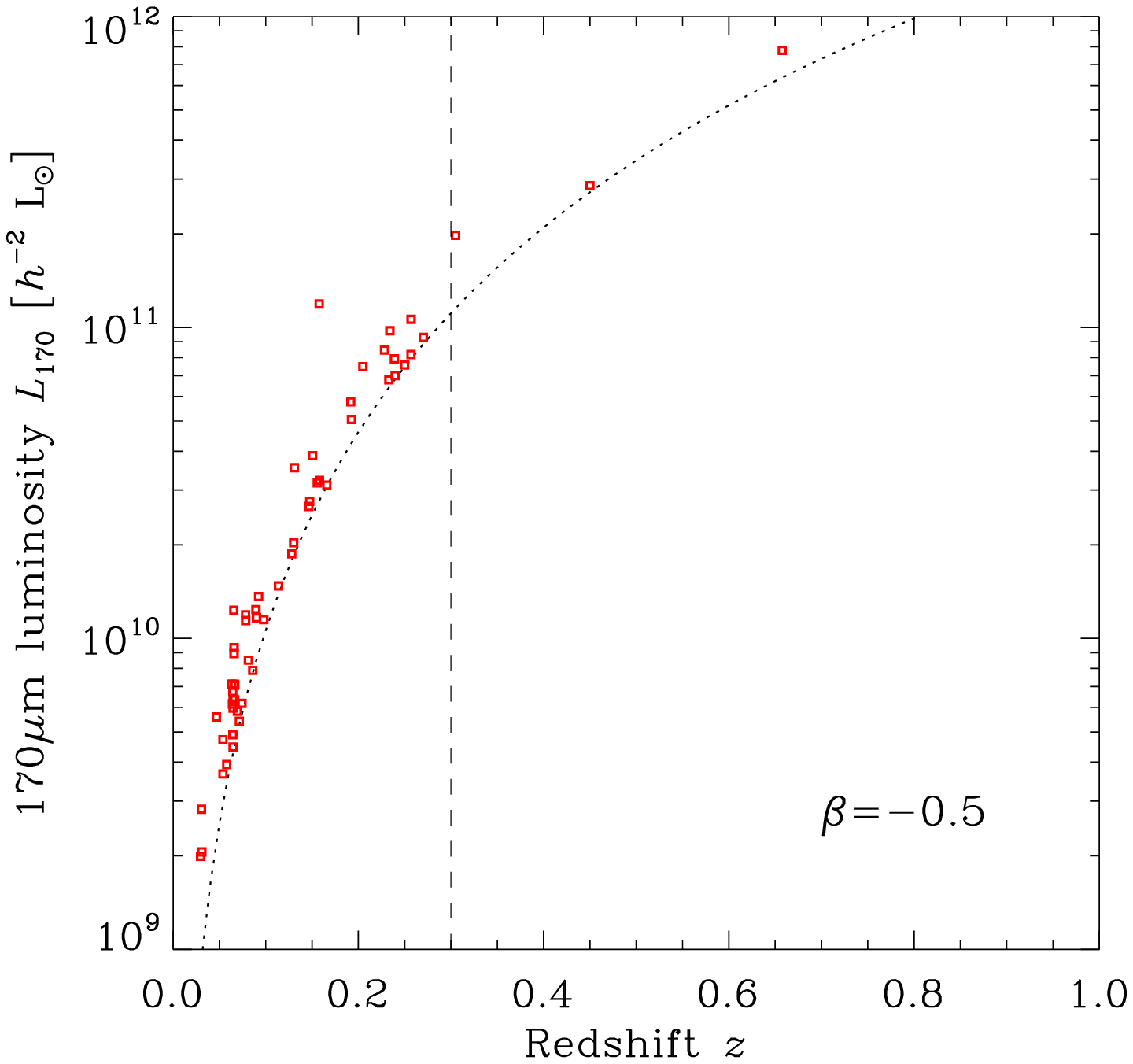}
\caption{The luminosity--redshift distribution of our redshift sample
taken from FIRBACK $170\,\mu$m survey.
The luminosity is that at emitted wavelength of 170$\mu$m.
The dotted curves represent the limiting luminosities corresponding to the
flux density detection limit of 195~mJy at each redshift, 
including the effect of the $K$-correction.
The power-law index of the spectral energy distribution (SED), $\beta$ is
1.0, 0.0, and $-0.5$ from left to right (see the main text). 
Vertical dashed lines show $z=0.3$, which is used to define our low-$z$
sample.
}\label{fig:lum_z}
\end{figure*}

The monochromatic luminosity at observed frequency $\nu_{\rm obs}$
is obtained by 
\begin{eqnarray}\label{eq:luminosity}
  {\cal L}_{\nu_{\rm em}} = {\cal L}_{(1+z)\nu_{\rm obs}}=
    \frac{4\pi d_{\rm L}(z)^2 S_{\nu_{\rm obs}}}{1+z} \;,
\end{eqnarray}
where ${\cal L}_\nu$ is an energy emitted per unit time at frequency $\nu$,
$d_{\rm L}(z)$ is the luminosity distance corresponding to a 
redshift $z$, and $\nu_{\rm obs}$ and $\nu_{\rm em}$ are observed and emitted
frequencies, respectively.
In order to estimate the luminosity function at $170\;\mu$m, 
the $K$-correction is required.

However, the amount of the $K$-correction may be uncertain, because 
the present sample is observed at one waveband.
If we assume a `cool' dust galaxy, the spectral energy distribution (SED) 
rises toward longer wavelengths, whilst it decreases if we adopt a starburst
SED \citep[see e.g.,][]{takeuchi01a,takeuchi01b,lagache03}.
{}To explore the effect of the $K$-correction, we use a power-law 
approximation with the form of 
\begin{eqnarray}
  {\cal L}_\nu \propto \nu^\beta\;.
\end{eqnarray}
As for $\beta$, we consider $\beta=1.0$ (starburst galaxies), 
$0.0$ (intermediate galaxies), and $-0.5$ (cool galaxies).
We adopt these values according to the phenomenologically constructed
model SEDs of \citet[][]{lagache03} (see their Figure~4).
Then, the luminosity at the observed frequency $\nu_{\rm obs}$ becomes
\begin{eqnarray}
  L_{\nu_{\rm obs}} &\equiv& \nu_{\rm obs}{\cal L}_{\nu_{\rm obs}} \nonumber \\
  &=& L_{\nu_{\rm em}}(1+z)^{-(\beta+1)} \nonumber \\
  &=& 4 \pi d_{\rm L}(z)^2 \nu_{\rm obs} S_{\nu_{\rm obs}}(1+z)^{-(\beta+1)}\;.
\end{eqnarray}
By the same manner, the limiting luminosity of a survey with flux density 
detection limit $S_{\rm \nu}^{\rm lim}$ depends on the SED via $\beta$,
\begin{eqnarray}\label{eq:llim}
  L_{\nu_{\rm obs}}^{\rm lim} = 
    4 \pi d_{\rm L}(z)^2 \nu_{\rm obs} S_{\nu_{\rm obs}}^{\rm lim}
    (1+z)^{-(\beta+1)}\;.
\end{eqnarray}
This is shown with the present sample in Figure~\ref{fig:lum_z}.
The luminosity $L_{170}$ is that at emitted wavelength of $170\;\mu$m, 
i.e., $L_{\rm em}$ measured at $170\;\mu$m.
As we see in the followings, this dependence of the limiting luminosity on
$\beta$ potentially affects the estimation of the LF.

\subsection{Estimation of the luminosity function}\label{subsec:lf}

We define the luminosity function as a number density of galaxies whose
luminosity lies between a logarithmic interval
$[\log L, \log L + d\log L]$:\footnote{We denote $\log x \equiv \log_{10} x$ 
and $\ln x \equiv \log_{e} x$, respectively.}
\begin{eqnarray}
  \phi (L) \equiv \frac{dn}{d \log L}\;.
\end{eqnarray}
In this work, we denote the luminosity at a certain frequency or wavelength 
as $L_\nu \equiv \nu {\cal L}_\nu$.

\subsubsection{Parametric estimation}

First we performed a parametric maximum likelihood estimation of the LF
\citep{sandage79}.
Note that we can do this analysis {\sl directly} 
on the data, being independent of the nonparametric result, i.e., this is
not a fitting to the nonparametric LF.
It is known that the 60-$\mu$m LF is well expressed by a function given by
\citet{saunders90} which is defined as
\begin{eqnarray}\label{eq:saunders}
  \phi (L) = \phi_* \left( \frac{L}{L_*} \right)^{1-\alpha}
    \exp \left[ -\frac{1}{2\sigma^2} 
    \log^2 \left(1+\frac{L}{L_*}\right)\right]\;
\end{eqnarray}
where $\log^2 x \equiv (\log x)^2$.
Since various LFs can be approximated by this functional form, we adopt
Equation~(\ref{eq:saunders}) in this work.

\begin{figure*}
\centering\includegraphics[angle=90,width=0.9\textwidth]{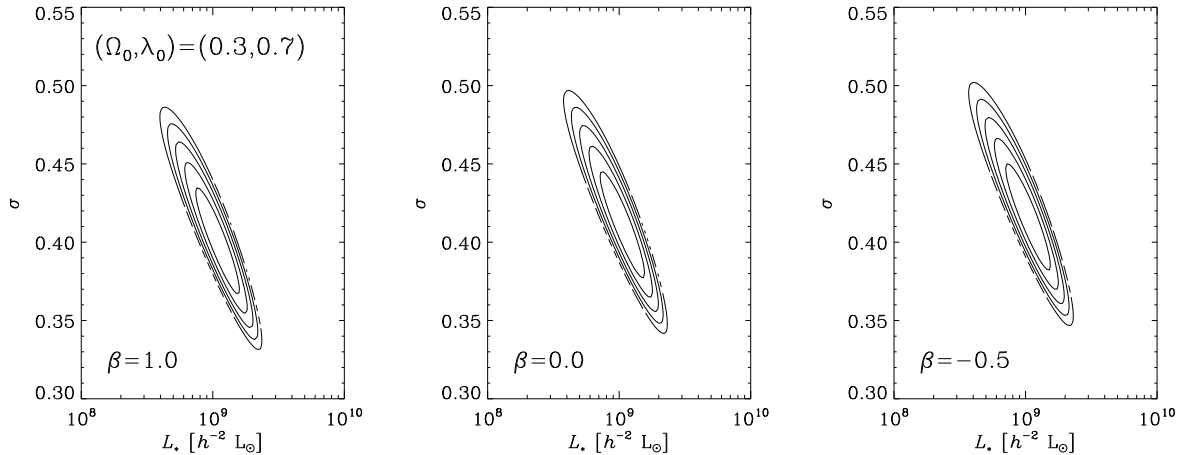}
\caption{The logarithmic likelihood $\ln {\cal M}$ for the parameter 
estimation of $L_*\;[L_\odot]$ and $\sigma$ when we fix $\alpha=1.25$.
The outermost contours indicate the 68~\% confidence level.
}\label{fig:lkh}
\end{figure*}

We use Equation~(\ref{eq:saunders}) for the parametric maximum likelihood
estimation.
Then the likelihood ${\cal M}$ is expressed as
\begin{eqnarray}\label{eq:lkh}
  &&{\cal M} (L_*,\alpha, \sigma |\{L_i,z_i\}_{i=1,\dots,N})
  = \prod_{i=1}^{N} \frac{\phi(L_i)}{\int_{\log L_{{\rm min},i}}^{\infty}
    \phi(L) d \log L} \nonumber\\
  &&= \prod_{i=1}^{N} \frac{\displaystyle
    \left(\frac{L_i}{L_*}\right)^{1-\alpha}
    \exp\left[ -\frac{1}{2\sigma^2}\log^2 \left(1+\frac{L_i}{L_*}\right)
    \right]}{\displaystyle
    \int_{\log L_{{\rm min},i}}^{\infty} \left(\frac{L}{L_*}\right)^{1-\alpha}
    \exp\left[ -\frac{1}{2\sigma^2}\log^2 \left(1+\frac{L}{L_*}\right)
    \right]d \log L} \nonumber \\
\end{eqnarray}
where 
\begin{eqnarray}\label{eq:lum_limit}
  L_{{\rm min},i} = L_{\nu_{\rm obs}}^{\rm lim}(z_i) \;.
\end{eqnarray}
Note that, in principle, the parametric estimation procedure is 
dependent on $\beta$.
We can obtain the parameters of the LF by maximizing Equation~(\ref{eq:lkh}) 
with respect to $L_*$, $\alpha$, and $\sigma$ 
\citep[][]{sandage79,saunders90,takeuchi03b}.

However, because of the small size of the present sample and relatively
narrow range of their luminosity, it is difficult to
put a reasonable constraint to the faint-end slope of the LF.
Hence instead, as we explain later, we assume a certain value for the 
faint-end slope $\alpha$.

\subsubsection{Nonparametric estimation}

We also estimate the LF nonparametrically via an improved version of 
the $C^-$ method of \citet{lyndenbell71}, implemented to have the density 
normalization \citep{choloniewski87}.
This method is a kind of maximum likelihood methods insensitive to
the density fluctuation.
This method and its extension are fully described and carefully examined by
\citet{takeuchi00}.\footnote{
We found that the other density-insensitive nonparametric estimators 
discussed in \citet{takeuchi00} were not very suitable for the present 
small sample analysis: Both of the methods of \citet{choloniewski86} and 
\citet{efstathiou88} need to divide the sample into small bins.
For the present sample (55 galaxies), we could not find stable solutions
for these estimators.
}

We note that the SED slope $\beta$ also affects the nonparametric 
estimation of the LF.
In the case of $C^-$ method, the definition of $C^-$ includes 
$L_{\rm \nu_{\rm obs}}^{\rm lim}$ \citep[see Figure~2 of ][]{takeuchi00}.
Thus, it will be important to explore the systematic effect introduced 
by $K$-correction.
The uncertainty (68~\% confidence limit) is estimated by the bootstrap 
resampling \citep[][]{takeuchi00}.
Additionally, we also estimated the LF and uncertainty including 
the observational measurement errors.
For the density normalization, we took into account the source extraction
completeness \citep[$\sim 90$~\%:][]{dole01} and the redshift measurement
(84~\%).

\section{Results}\label{sec:results}

\subsection{Parametric result}\label{subsubsec:param}

We fixed the faint-end slope of the FIRBACK 170-$\mu$m LF to be $1.25$.
This is very close to that of the total 60-$\mu$m LF of the 
{\sl IRAS} PSC$z$ 
sample, and the same as that of its `cool' subsample \citep{takeuchi03b}.

We obtained $L_*=1.1 \times 10^9h^{-2}\;[L_\odot]$ 
(with a 68-\% confidence range of $3.0 \times 10^8h^{-2}
\mbox{--}2.3 \times 10^9 h^{-2}\; [L_\odot]$)
and $\sigma=0.41$ (with a 68-\% confidence range of $0.35\mbox{--}0.50)$.
The likelihood contours are shown in Figure~\ref{fig:lkh}.
The outermost contours indicate $\Delta \ln {\cal M} \equiv \ln {\cal M} -
\ln {\cal M}_{\rm max} = -0.5$, corresponding to the 68-\% confidence limit.
In Figure~\ref{fig:lkh}, we present $\ln {\cal M}$ defined by 
Equation~(\ref{eq:lkh}), as a function of $(L_*, \sigma)$.
These two parameters are rather strongly dependent with one another, and as a
result, the contour is elongated along with the diagonal direction in each
panel.
The density normalization was $\phi_*=(1.0 \pm 0.4) \times 10^{-1} h^3 \;
[\mbox{Mpc}^{-3}]$.
As seen in Figure~\ref{fig:lkh}, the result is almost independent of the 
assumed $\beta$.
The result is also found to be quite robust against the value of $\alpha$
in a plausible range of $\alpha=1.1\mbox{--}1.3$.

\citet{takeuchi03b} presented the parameters for the LF of the {\sl IRAS} 
PSC$z$ sample.
The parameters for the LF of the whole sample are $(\alpha, L_*, \sigma)=
(1.23, 4.34\times 10^{8}h^{-2}\;[L_\odot], 0.724)$.
Clearly, the parameters for the 170-$\mu$m sample are different from these
values.
Particularly, $\sigma$ which determines the steepness of the bright end 
is significantly smaller than the total {\sl IRAS} LF.
{}From the above likelihood analysis, $\sigma=0.724$ was rejected with 
a confidence level of more than 99.9~\%, even for the small number of galaxies.
That is, the bright end of the present LF declines more steeply than
that of the total {\sl IRAS} LF.

\subsection{Nonparametric result}

\begin{figure*}
\centering\includegraphics[width=0.45\textwidth]{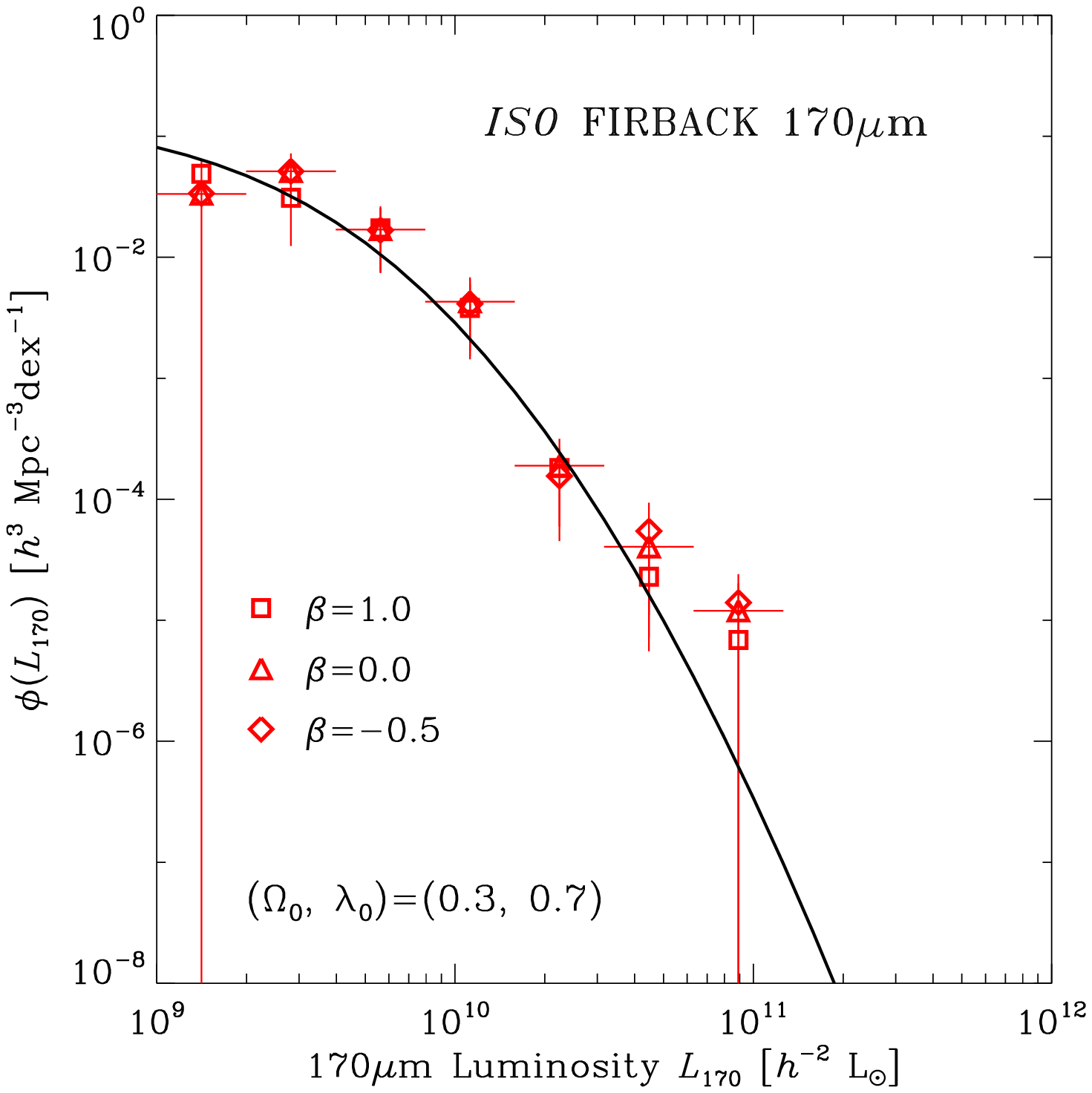}
\centering\includegraphics[width=0.45\textwidth]{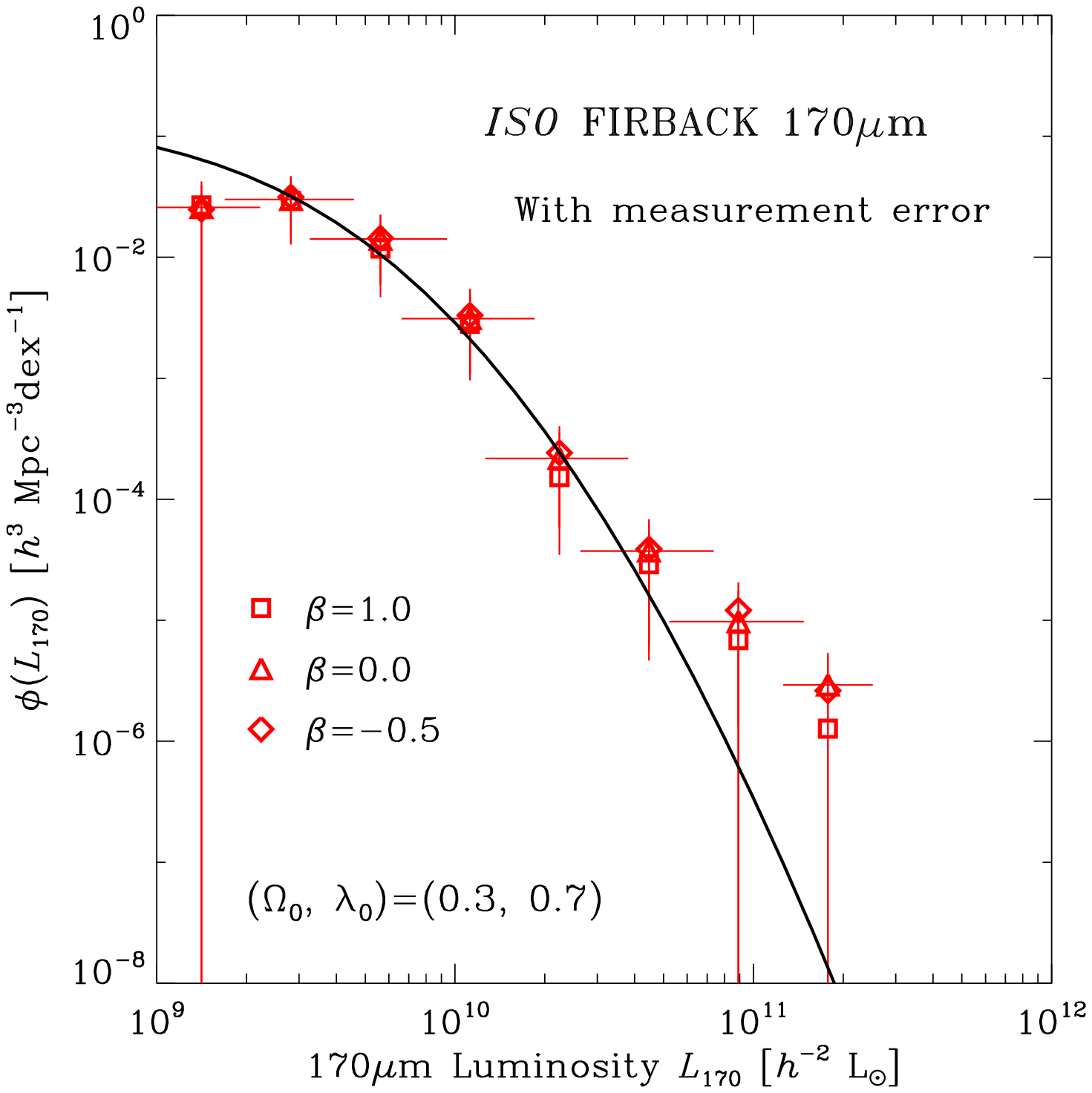}
\caption{The local luminosity function of {\sl ISO} FIRBACK 170~$\mu$m galaxy 
sample.
Solid curves show the parametric form estimated from Equation~(\ref{eq:lkh}).
Symbols represent the $C^-$ nonparametric LFs, respectively.
Open squares, open triangles, and open diamonds are the LFs adopting 
$\beta=1.0$, 0.0, and $-0.5$, respectively.
Vertical error bars are {68-\% confidence ranges.}
Left panel shows the LF from the original data, while the right panel shows
the LF from the data convolved with observational measurement errors.
In the left panel, horizontal bars simply represent the bin width (0.3~dex).
In the right panel, in contrast, horizontal bars are the convolution of 
the bin width and the luminosity uncertainty introduced by the photometric
error.
For visual simplicity, we put horizontal bars only on the case of 
$\beta=0.0$ in each panel.
}\label{fig:lf}
\end{figure*}

Nonparametric LF estimates are presented in Figure~\ref{fig:lf}.
Symbols are the LFs obtained by the $C^-$-method.
Vertical error bars show the 68~\% uncertainty, obtained by bootstrap 
resampling of $10^4$ times.
Open squares, open triangles, and open diamonds represent $\beta=1.0$, 
0.0, and $-0.5$, respectively.
Solid lines are the analytic expression of the LF with the parameters
we obtained in Section~\ref{subsubsec:param}.
We present the numerical tables of our $C^-$ LF estimates also in Appendix~A.

The left panel is the LF estimates from the original data themselves, while
the right panel is the ones from the data convolved with measurement errors.
We performed the error convolution procedure assuming a Gaussian distribution
for the measurement errors in flux density, with quoted flux density errors
as standard deviations for each data.
This procedure slightly blurs the LFs horizontally.
As a result, we have an additional bin at the highest luminosity, and
the 68~\% uncertainty levels are broadened.
However, the effect is rather small, and the LF estimates are quite robust
except for the highest-luminosity unstable bins.

We see that different values for $\beta$ do not affect the result 
very strongly, but at the highest-luminosity bins, systematic
effects are relatively large, a factor of $0.5\mbox{--}0.7\;\mbox{dex}$.
In contrast to the large effect of $\beta$ we found in 
Figure~\ref{fig:lum_z}, the LF estimates are quite robust against $\beta$.
Hence, hereafter, we do not show all the results with respect to $\beta$,
but only restrict ourselves to the results with $\beta=0.0$ without  
loss of generality.

\begin{figure}
\centering\includegraphics[width=0.45\textwidth]{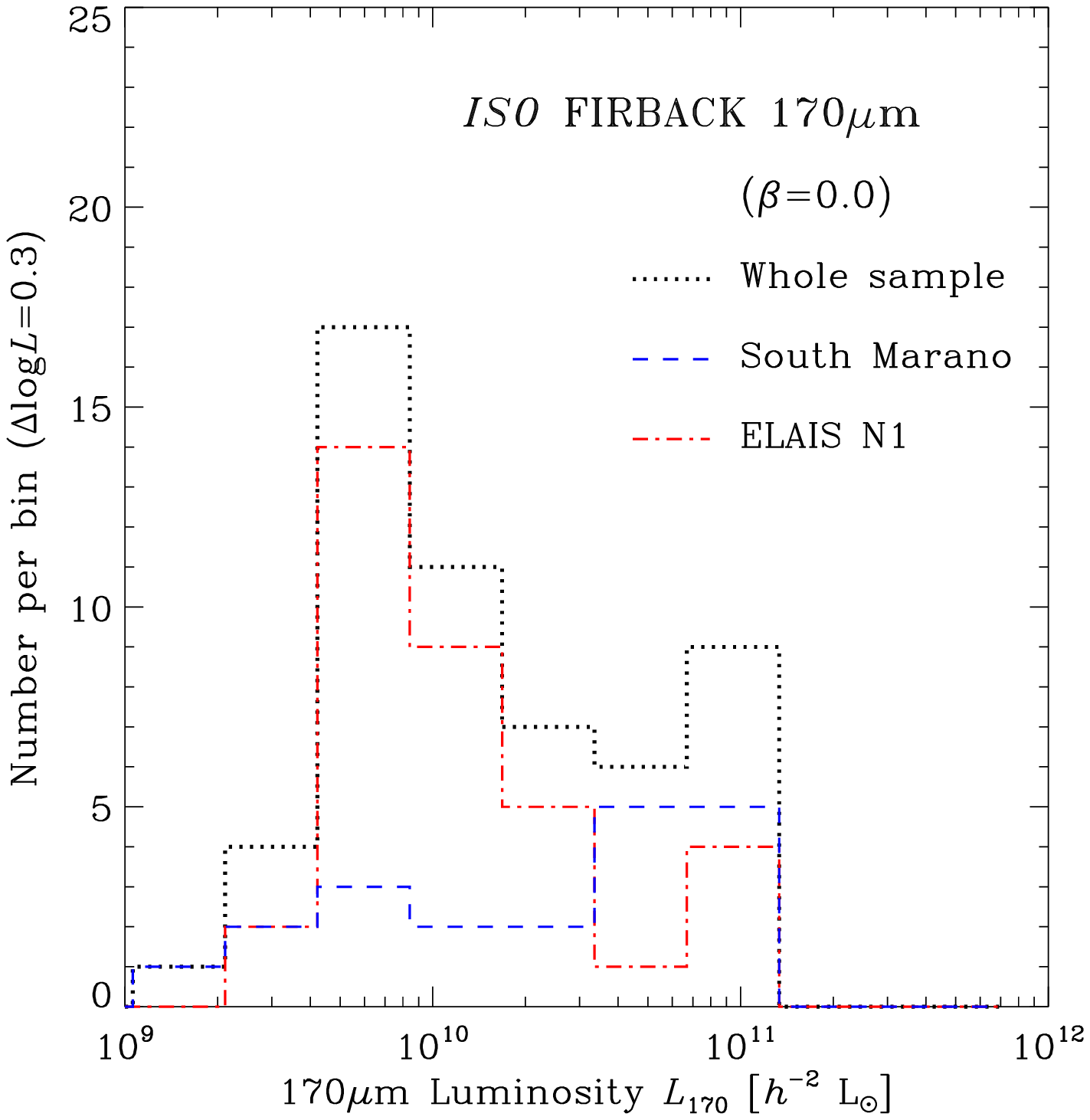}
\caption{The luminosity distribution of the FIRBACK 170~$\mu$m low-$z$
galaxy sample.
The dotted histogram is the luminosity distribution of the whole sample.
It has an excess at the highest luminosity bin.
The dashed and dot-dashed histograms are those of the subsamples in 
South Marano field and ELAIS N1 field, respectively.
}\label{fig:ldist}
\end{figure}

\begin{figure*}
\centering\includegraphics[width=0.7\textwidth]{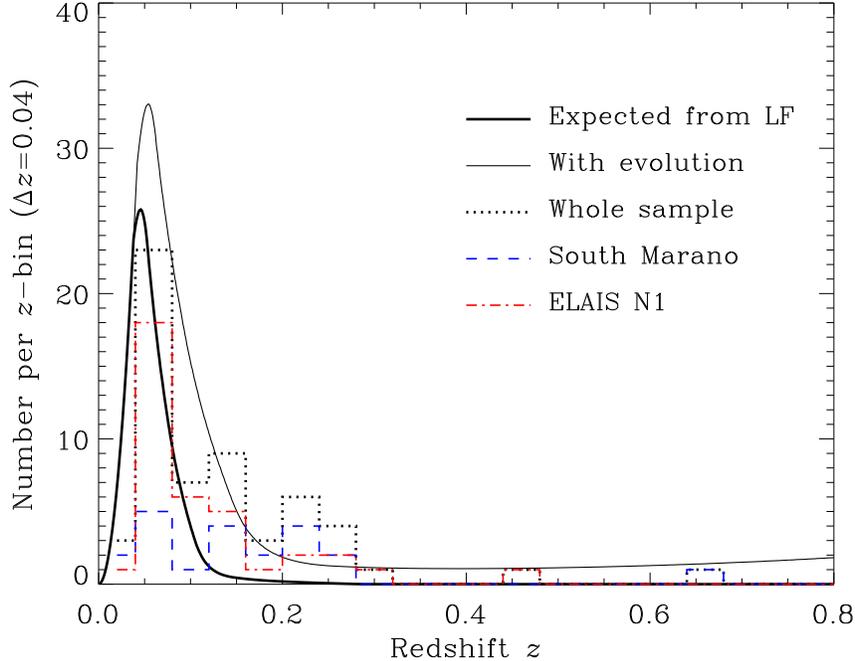}
\caption{The redshift distribution of our flux-limited sample.
The dotted histogram is the distribution of the whole sample,
the dashed histogram shows the the subsample in South Marano field, 
and dot-dashed one presents the subsample in ELAIS N1 field.
These histograms present the number of galaxies in each bin ($\Delta z=0.04$).
The thick solid curve is the expected distribution of galaxies 
calculated from the nonparametric local LF (Figure~\ref{fig:lf}).
The thin solid curve is the expected distribution for the
pure luminosity evolution with $L(z) \propto (1+z)^{5.0}$.
}\label{fig:zdist}
\end{figure*}

We should note the upward deviation of the $C^-$ LF from the analytic 
result.
Although the error bar is large, the $C^-$ estimates 
(open symbols in Figure~\ref{fig:lf}) are about an order of magnitude larger
than that of the analytic value.
It is worth examining if this deviation is real or merely an artifact of the
poor statistics.
{}To see this, we show the luminosity distribution of our present 
sample in Figure~\ref{fig:ldist}.
Recall that this sample consists of galaxies only at $z < 0.3$.
We see an excess at the highest luminosity bin.
Examining the subsamples, we also find a similar excess in the ELAIS N1 field.
We, however, also see that the sample in South Marano field is interesting:
the galaxies in this subsample are biased toward higher luminosities.
The superposition of these effects makes the brightest end of the 
nonparametric LF deviated from the analytic one.

Why does the analytic maximum likelihood not reflect this sample property?
We see that, though there is an excess at $L_{170} \simeq 10^{11} h\;L_\odot$,
the majority of the sample is located around $10^{10}h\;L_\odot$ (the 
peak in Figure~\ref{fig:ldist}).
Hence, the parameter estimation is practically controlled by the sample
around $L_*$, and the bright galaxies could hardly give a strong effect to 
the final estimation.
In addition, the peak is dominated by the sample from ELAIS N1 field, and
the peculiar luminosity distribution of South Marano field affected 
the result little.
We, hence, should not overly rely on the analytic result and parameters, 
but rather we should use the nonparametric $C^-$ LF directly.
These are small sample effects, and we should wait for the larger
sample to clarify the detailed shape of the LF.

We show the redshift distribution of the present sample in 
Figure~\ref{fig:zdist}.
The histograms show the distribution of the present data.
The thick solid curve is the expected redshift distribution of galaxies 
calculated from the nonparametric local LF we have obtained.
This is calculated as follows
\begin{eqnarray}\label{eq:dndz}
  \frac{dN}{dz}= \Omega_{\rm survey} \int_{\log L_{\rm min}(z)}^{\infty}
    \phi(L_{170})\frac{d^2V}{d\Omega dz} d\log L_{170} \;,
\end{eqnarray}
where $L_{\rm min}(z)$ is the minimum detectable luminosity 
$L_{{\nu}_{\rm obs}}^{\rm lim}(z)$ [see, Eq.~(\ref{eq:lum_limit})], and
\begin{eqnarray}\label{eq:comoving}
  \frac{d^2V}{d\Omega dz} = \frac{c}{H_0}
    \frac{d_{\rm L}(z)^2}{(1+z)^2\sqrt{\Omega_0 (1+z)^3 + \lambda_0}}
\end{eqnarray}
for the flat lambda-dominated universe \citep[see, e.g.,][]{peebles93},
i.e., $(d^2V/d \Omega dz)dz$ is the comoving volume between $[z,z+dz]$ per 
unit solid angle.
This curve is calculated to apply to the solid angle of the survey area
by multiplying $\Omega_{\rm survey}=8.4 \times 10^{-4}\;\mbox{sr}$.
We used the nonparametric LF for Figure~\ref{fig:zdist} because the exact
shape of the bright end of the LF is crucial to examine the tail of the 
redshift distribution of the source toward higher $z$, and as already
discussed, the analytic function underproduces galaxies with 
$L_{170} \ga 3 \times 10^{10}\;L_\odot$.
We find a density excess at $z \simeq 0.2$ in 
Figure~\ref{fig:zdist}.\footnote{If we use the classical 
$1/V_{\rm max}$-method \citep{schmidt68,eales93}, the estimator is
affected by this bump and results in a (fake) flatter bright end of the LF.
This is a well-known drawback of the $1/V_{\rm max}$-method
\citep[][and references therein]{takeuchi00}, and we must
be very careful about its usage.
We address this problem in Appendix~B.}
However, apart from the bump, both fields have long tails toward higher 
redshifts than expected from the local LF.
It may suggest the existence of galaxy evolution.
We explore the effect of the evolution in Section~\ref{subsec:evolution}

\section{Discussion}

\begin{figure}
\centering\includegraphics[width=0.45\textwidth]{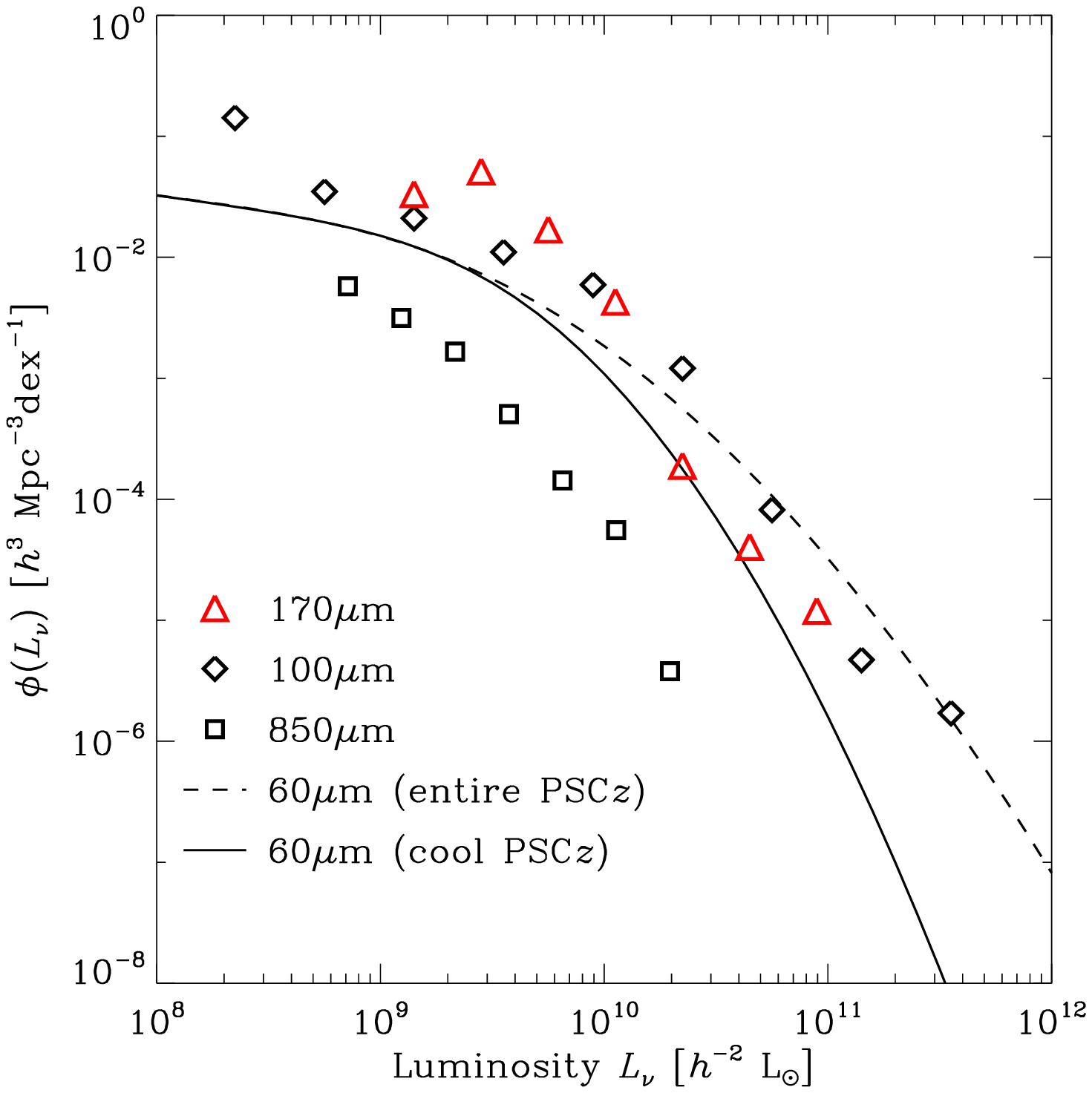}
\caption{Comparison of the 170-$\mu$m LF with those at other FIR wavelengths.
Open triangles are our 170-$\mu$m LF ($\beta=0.0$), 
open diamonds represent the 100-$\mu$m LF of \citet{soifer91}, and 
open squares represent the 850-$\mu$m LF of \citet{dunne00}.
The dashed curve is the analytic form of the 60-$\mu$m LF of 
the {\sl IRAS} PSC$z$ entire sample, and the solid curve is the LF
of the cool subsample of the {\sl IRAS} PSC$z$ galaxies
\citep{takeuchi03b}.
The 850-$\mu$m LF is horizontally shifted by a factor of 100 for 
display purposes.
}\label{fig:lf_compare}
\end{figure}

\subsection{Shape of the $170\;\mu$m LF}

The overall shape of the 170-$\mu$m LF of the FIRBACK galaxies is 
different from the {\sl IRAS} 60-$\mu$m LF.
As we discussed in Section~\ref{subsubsec:param}, the parameters of the 
analytic LF suggest a steeper slope for the bright end of the LF.
Although the nonparametric LF revealed that the brightest part of the LF 
is significantly higher than that of the analytic LF, the 170-$\mu$m LF
decreases more rapidly from the knee of the LF to the bright end than
the {\sl IRAS} 60-$\mu$m LF.
Here we compare the 170-$\mu$m LF with other FIR LFs obtained to date.

\citet{takeuchi03b} divided the {\sl IRAS} sample into two categories, 
warm and cool subsamples, using the flux density ratio criterion of 
$S_{100}/S_{60} = 2.1$.
The parameters of the cool galaxies are 
$(\alpha,L_*, \sigma)=(1.25,9.55\times 10^{9}h^{-2}\;[L_\odot], 0.50)$.
The $\alpha$ is almost the same as that of the total {\sl IRAS} LF
within the quoted error, but $L_*$ and $\sigma$ are much closer to those of
our present LF.
The resemblance of our LF to the LF of cool {\sl IRAS} galaxies may be 
reasonable, since the present sample is selected at $170\;\mu$m band, 
where we can detect cooler galaxies more effectively than $60\;\mu$m.
\citet{soifer91} derived a LF at $100\;\mu$m from {\sl IRAS}
galaxy sample.
Their 100-$\mu$m LF has a bright end slightly steeper than 60-$\mu$m LF does,
but the slope is still flatter than our LF.
\citet{dunne00} constructed a LF at $850\;\mu$m based on 60-$\mu$m selected
{\sl IRAS} galaxy sample.
Although the dynamic range is small, the overall shape of their 850-$\mu$m
LF is similar to our LF.
Actually, \citet{dunne00} fitted their LF with the Schechter function which 
has a very steep decline at the bright end \citep{schechter76}.\footnote{
However, we should keep in mind that the sample of \citet{dunne00} is selected
at $60\;\mu$m.
As these authors discussed, a significant fraction of galaxies with cold dust
may be missed in the sample, and consequently, it might be possible that 
their LF shape is not representative of cool/cold dust galaxies.
}

In summary, the 170-$\mu$m LF has a shape similar to those of galaxy
sample with cool dust emission.
It is also interesting to note that the 170-$\mu$m LF has the highest
normalization among known FIR LFs.
We summarize the comparison in Figure~\ref{fig:lf_compare}.
However, we must keep in mind the large uncertainty of the estimates,
and further observational exploration is definitely required.

\subsection{Evolution}\label{subsec:evolution}

\subsubsection{Pure luminosity evolution assumption}

Most of the galaxies in our redshift sample are at $z<0.3$.
Hence, we should estimate the evolution of the LF under a certain assumption.
We adopt a pure luminosity evolution (PLE).
Recent {\sl Spitzer} observations indicated that the actual evolution of 
IR galaxies is described as a strong evolution in luminosity, with a slight
evolution in density \citep[e.g.,][]{perez_gonzalez05,lefloch05}, while
their studies are based on the {\sl Spitzer} $24\;\mu$m band.
Hence, the PLE is not a bad choice as a first approximation.

Adopting the PLE, the strength of the evolution can be estimated via a radial 
density distribution of galaxies \citep{saunders90}.
In the case of the PLE, the LF at $z$ is expressed by the evolution strength
$f(z)$ as
\begin{eqnarray}
  \phi(L,z) = \phi_0\left[\frac{L}{f(z)}\right] \;,
\end{eqnarray}
where $\phi_0(L)$ is the local functional form of the LF.
The PLE assumes that this form remains unchanged and only shifts along the 
luminosity axis.
{}To define $\phi_0(L_{170})$, we examined the 170-$\mu$m LF for a sample
at $z < 0.15$ (38 galaxies).
Though the uncertainty is large, we did not observe a significant change 
of the LF in its shape, i.e., the PLE assumption might be approximately 
valid.
Hence, we can use the shape of the LF of the sample with $z < 0.3$ as 
$\phi_0(L_{170})$.
For simplicity, we consider a power-law form for $f(z)$:
\begin{eqnarray}\label{eq:evol_power}
  f(z)=(1+z)^{Q} \;,
\end{eqnarray}
but this form is supported by recent {\sl Spitzer} observations at a wide 
redshift range of $z=0\mbox{--}1$ \citep{perez_gonzalez05,lefloch05}.

\subsubsection{Estimation of evolution via radial density}

Assuming that the LF is separable for $L$ and $z$ as $\phi(L,z)=n(z)p(L)$, 
the likelihood is written as
\begin{eqnarray}\label{eq:rad_lkh}
  {\cal M} (Q|\{L_i,z_i\}_{i=1,\dots,N})
  = \prod_{i=1}^{N} \frac{\phi(L_i,z_i)}{\displaystyle
    \int_{z_{\rm min}}^{z_{{\rm max},i}}
    \phi(L_i,z) \frac{d^2V}{d\Omega dz} dz}
\end{eqnarray}
where $N$ is the total sample size, $z_{\rm min}$ is the lowest redshift
which we used in the analysis,
$z_{{\rm max},i}$ is the maximum redshift to which $i$th galaxy can be 
detected, and $(d^2V/d \Omega dz)dz$ is again the differential 
comoving volume [Eq.~(\ref{eq:comoving})].

By maximizing Equation~(\ref{eq:rad_lkh}) with respect to $Q$, we can have
a maximum likelihood estimate.
We performed this procedure with the nonparametric LF, because as we already
mentioned, the analytic form underestimates the bright end of the LF, and it
would lead to a serious overestimation of the evolution strength.
We obtained $Q=5.0^{+2.5}_{-0.5}$.
The expected redshift distribution with this evolution is obtained by 
a similar manner to Equation~(\ref{eq:dndz}) as
\begin{eqnarray}
  \frac{dN}{dz}= \Omega_{\rm survey} \int_{\log L_{\rm min}(z)}^{\infty}
    \phi_0 \left[\frac{L_{170}}{(1+z)^Q}\right]
    \frac{d^2V}{d\Omega dz} d\log L_{170} \;,
\end{eqnarray}
and shown in Figure~\ref{fig:zdist} (the thin solid line).
The agreement with the data and the expected value is much improved.

Although the uncertainty is very large because of the limited sample size,
this is similar to the value obtained by recent {\sl Spitzer} 24-$\mu$m
observations, $Q \simeq 4$ \citep{perez_gonzalez05,lefloch05}.
These authors also found a weak evolution in the galaxy number density,
but for the present sample, it is impossible to explore this effect.
If confirmed, the similarity between the strength of the galaxy evolution 
in the MIR (12-$\mu$m in the rest frame) and FIR will provide us an 
important clue to the physics of dusty star formation in galaxies.

\subsection{FIR luminosity density and obscured star formation density in 
the Local Universe}

\subsubsection{The local 170-$\mu$m and total IR luminosity density}

\begin{figure}
\centering\includegraphics[width=0.45\textwidth]{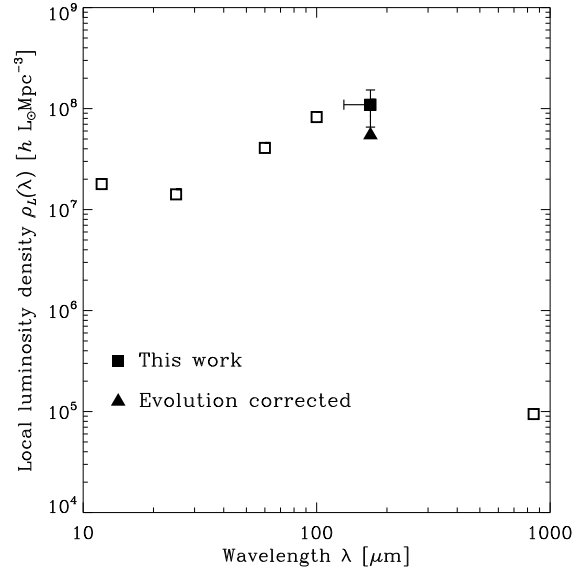}
\caption{The MIR and FIR luminosity densities in the Local Universe.
The filled square represent the 170-$\mu$m luminosity density, 
$\rho_L(170\mu\mbox{m})$, obtained by a direct integration of our LF.
The vertical error bar shows the total uncertainty related to the LF.
The horizontal bar represents the redshift range of our local sample 
($0 \le z < 0.3)$.
The filled triangle is the corrected $\rho_L(170\mu\mbox{m})$ by assuming
a pure luminosity evolution $L(z) \propto (1+z)^5$.
Open squares present the luminosity densities at 12, 25, 60, 100, and 
$850\;\mu$m calculated based on the literatures (see text).
}\label{fig:local_sed}
\end{figure}

The luminosity density in a cosmic volume provides various information of
the energy distribution in the Universe.
Especially, comparison between the radiative energy directly emitted from 
stars and that re-emitted from dust is one of the key quantities to 
understand the fraction of hidden star formation.
In this subsection, we discuss the local luminosity density at $170\;\mu$m, 
$\rho_L(170\mu\mbox{m})\; [L_\odot \mbox{Mpc}^{-3}]$
and consider the integrated SED of the Local Universe.

In principle, it is straightforward to obtain 
$\rho_L(170\mu\mbox{m})$ from the LF: we simply
integrate the first-order moment of the LF, 
$L_{170}\phi(L_{170})$, over the whole possible range
of luminosity.
Since the luminosity range of the 170-$\mu$m LF is limited to 
$10^9 \mbox{--}10^{12}\;L_\odot$, we extrapolated the faint end without
observed data.
We have done it by using the analytic form (Eq.~\ref{eq:saunders}) with
various $\alpha$.
However, as far as $\alpha < 2.0$, the integration of $L_{170}\phi(L_{170})$
converges and the faint end does not contribute to the total integration
significantly.
For the infrared (IR) galaxies, previous studies suggest $\alpha < 2.0$ 
\citep[e.g.,][and references therein]{saunders90,takeuchi03b}, and the 
$\rho_L(170\mu\mbox{m})$ is little affected by $\alpha$.
By the same reason, the lower and upper bounds of the integration do not 
affect the result.
We chose $10^7h^{-2} [L_\odot]$ as the lowest luminosity, and use the highest
luminosity bin as the upper bound.
Thus, we obtained $\rho_L(170\mu\mbox{m}) = (1.1^{+0.5}_{-0.4})\times 10^8 h 
[L_\odot \mbox{Mpc}^{-3}]$.
The final uncertainty of $\rho_L(170\mu\mbox{m})$ is 
dominated by the statistical uncertainty of the nonparametric LF in each bin.

We, however, must recall that the present FIRBACK sample is not exactly
`local', i.e., it consists of galaxies $0 \le z < 0.3$.
As seen in the previous subsection, it may be plausible that this result
is affected by the strong galaxy evolution.
If we adopt a PLE $L(z) \propto (1+z)^{5.0}$, $\rho_L(170\mu\mbox{m})$ should
be enhanced by a factor of 2.1 on average in this redshift range.
By correcting the evolution, we have $\rho_L(170\mu\mbox{m})=5.20\times 10^7 h 
[L_\odot \mbox{Mpc}^{-3}]$. 

We plot the result in Figure~\ref{fig:local_sed}.
The filled square represents the 170-$\mu$m luminosity density obtained by 
a direct integration of our LF.
The horizontal bar indicates that the redshift range of our low-$z$ sample 
is $0 \le z < 0.3$.
The filled triangle shows the evolution-corrected value.
Then, we consider the SED of the luminosity density in the Local Universe.
We do similar exercises to calculate $\rho_L(\lambda)$ at various IR 
wavelengths.
At MIR, \citet{fang98} and \citet{shupe98} provided the LFs
at {\sl IRAS} 12 and $25\;mu$m bands, respectively.
\citet{fang98} tabulated their nonparametric LF at approximately the same
luminosity range as we adopt in this work ($10^7 L_\odot \mbox{--} 
10^{12}L_\odot$).
We simply summed it up with multiplying the luminosity and obtained
$\rho_L(12 \mu\mbox{m})=1.79 \times 10^7 h \;[L_\odot \mbox{Mpc}^{-1}]$.
\citet{shupe98} provided an analytic fit for their 25-$\mu$m LF.
By their analytic function, we got $\rho_L(25 \mu\mbox{m})=1.42 \times 10^7
h \;[L_\odot \mbox{Mpc}^{-1}]$.
At $60\;\mu$m, from \citet{takeuchi03b}, we obtained 
$\rho_L(60 \mu\mbox{m})=4.08 \times 10^7 h \;[L_\odot \mbox{Mpc}^{-1}]$.
\citet{soifer91} presented the LFs at all the {\sl IRAS} bands.
Using their 100-$\mu$m nonparametric LF, we re-calculated the luminosity
density to obtain 
$\rho_L(100 \mu\mbox{m}) = 8.25\times 10^7 h \;[L_\odot \mbox{Mpc}^{-1}]$.
Lastly, we used the Schechter function fit provided by \citet{dunne00} to
have $\rho_L(850 \mu\mbox{m})$.
Since their faint-end slope is very steep ($\alpha = 2.12$), the integration
is dependent on the adopted lowest luminosity in this case.
We coherently integrated the Schechter function in the same range as the 
other bands.
We found  $\rho_L(850 \mu\mbox{m})=1.00 \times 10^5 h 
\;[L_\odot \mbox{Mpc}^{-1}]$.

We plot these luminosity densities in Figure~\ref{fig:local_sed} 
(open squares).
The overall peak of the SED of the luminosity density seems to lie at 
$\lambda \ga 100\;\mu$m.
Further, if the suggested strong evolution is true, the local SED peak is 
restricted to be at $100 \ga \lambda \ga 170\;\mu$m.
Even though the evolution effect significantly reduces the local 
$\rho_L(170\mu\mbox{m})$ value, it still considerably contributes to 
the total FIR luminosity density.

\begin{table}
\centering
  \caption{The SED of the luminosity density in the Local Universe.}
\label{tab:local_sed}
  \begin{tabular}{ccc}
   \hline
    $\lambda$ & 
      $\rho_L(\lambda)$ & 
      $\rho_L(\lambda)/\rho_L(170\,\mu\mbox{m})$ \\
      $[\mu\mbox{m}]$ & 
      $h\;[L_\odot\mbox{Mpc}^{-3}]$ \\
   \hline
      12  & $1.79 \times 10^{7}$ & $3.3\times 10^{-1}$ \\
      25  & $1.42 \times 10^{7}$ & $2.6\times 10^{-1}$ \\
      60  & $4.08 \times 10^{7}$ & $7.5\times 10^{-1}$ \\
      100 & $8.25 \times 10^{7}$ & $1.5$ \\
      170 & $5.20 \times 10^{7}$ $^{\mathrm{a}}$ & $1$ \\
      850 & $9.45 \times 10^{4}$ $^{\mathrm{b}}$ & $1.7\times 10^{-3}$ \\
   \hline
  \end{tabular}
  \begin{list}{}{}
    \item[$^{\mathrm{a}}$] The effect of the evolution is corrected assuming
      a pure luminosity evolution with $L(z)\propto (1+z)^Q$ ($Q=5.0$).
    \item[$^{\mathrm{b}}$] This value is calculated by integrating 
      a Schechter function presented by \citet{dunne00}, over the luminosity
      range of $L_{850}=10^7 L_\odot \mbox{--}10^{13}L_\odot$.
  \end{list}
\end{table}

{}To have a crude estimate of the total IR luminosity density 
(we call it $\rho_L(\mbox{dust})$), we logarithmically interpolate and 
extrapolate between the SED data points in units of $[\mbox{erg\,s}^{-1},
\mbox{Hz}^{-1}\mbox{Mpc}^{-3}]$, and integrate it over the range of 
$8\mbox{--}1000\;\mu$m to match the conventional definition of the 
total IR (TIR) luminosity \citep[see, e.g., ][]{dale01,dale02,takeuchi05a}.
If we do not assume the evolution, we have 
$\rho_L(\mbox{dust}) = 1.4 \times 10^{8}h\;[L_\odot \mbox{Mpc}^{-3}]$, 
and with the evolution, $\rho_L(\mbox{dust})=1.1 \times 10^{8}h\;
[L_\odot \mbox{Mpc}^{-3}]$.
\citet{takeuchi05c} estimated the $\rho_L(\mbox{dust})$ under the assumption
of a constant ratio of $\rho_L(\mbox{dust})/\rho_L(60\mu\mbox{m})=2.5$
\citep[see, e.g.,][]{takeuchi05a}.
They found $\rho_L(\mbox{dust})=1.02 \times 10^{8}h\;
[L_\odot \mbox{Mpc}^{-3}]$.
Our $\rho_L(\mbox{dust})$ values are in a very good agreement with that,
especially for the case with evolution.

For the interpretation of the IR luminosity density, it should be worth 
mentioning that there is a possibly high contamination by 
active galactic nuclei (AGN) at 12 and 24~$\mu$m.
However, by integrating over the energy density from these wevelengths,
we find that the contribution from 12 and 24~$\mu$m bands to 
$\rho_L(\mbox{dust})$ is less than 10~\%.
Hence, if all the energy from 12 and 24~$\mu$m were from the AGN, the
effect of AGNs would not change the physical interpretation of 
$\rho_L(\mbox{dust})$ significantly.

\subsubsection{The obscured star formation density in the Local Universe}

The ratio between the energy from young stars directly observed at UV
and that reprocessed by dust and observed at IR in the cosmic history
has long been a matter of debate.
Before closing the discussion, we consider the star formation rate density
in the Local Universe obscured by dust.
We use the value with evolutionary correction in the rest of this work.
{}To get values without this evolutionary correction, we may simply
substitute the former value for $\rho_L(\mbox{dust})$.

For the conversion from $\rho_L(\mbox{dust})$ to the cosmic star formation
rate (SFR) density related to dust, $\rho_{\rm SFR}(\mbox{dust})$, we can
use several methods.
\citet{kennicutt98} presented a famous formula between the SFR and the
dust luminosity, $L_{\rm dust}\;[L_\odot]$,
\begin{eqnarray}\label{eq:kennicutt}
  \mbox{SFR} \; [M_\odot \,\mbox{yr}^{-1}] = 1.72 \times 10^{-10}\;
    L_{\rm dust}\;[L_\odot] \;,
\end{eqnarray}
which is valid for {\sl starburst} galaxies with a burst younger than 
$10^8$~yr.
Adopting this formula to $\rho_L(\mbox{dust})$, we obtained
$\rho_{\rm SFR}(\mbox{dust}) = 1.89\, h \times 10^{-2} 
\; [M_\odot \,\mbox{yr}^{-1}\mbox{Mpc}^{-3}]$.
However, as mentioned by \citet{kennicutt98} himself, this formula is not
valid for more quiescent, normal galaxies.
Since our SED of the Local Universe is similar to a kind of cool galaxies,
we should carefully treat the effect of the heating radiation from old stars.
\citet{hirashita03} found that about 40~\% of the dust heating in the nearby
galaxies comes from stars older than $10^8$~yr.
If we apply the correction of old stellar population to 
Equation~(\ref{eq:kennicutt}), we obtain $\rho_{\rm SFR}(\mbox{dust}) = 
1.14\, h \times 10^{-2} \; [M_\odot \,\mbox{yr}^{-1}\mbox{Mpc}^{-3}]$.
\citet{bell03} also presented a similar correction factor 
($0.32 \pm 0.16$ for galaxies with $L_{\rm dust} \le 10^{11}L_\odot$ and 
$0.09 \pm 0.05$ for those with $L_{\rm dust} > 10^{11}L_\odot$) 
for the contribution of old stars.
Based on a more theoretical point of view, \citet{takeuchi05c} also obtained 
an appropriate formula including the correction, which can be written as
\begin{eqnarray}\label{eq:tbb}
  \mbox{SFR} \; [M_\odot \,\mbox{yr}^{-1}] = 
    1.07 \times 10^{-10} \; L_{\rm dust}\;[L_\odot] \;.
\end{eqnarray}
Adopting Equation~(\ref{eq:tbb}), we obtain 
$\rho_{\rm SFR}(\mbox{dust}) = 1.17\, h \times 10^{-2} 
  \; [M_\odot \,\mbox{yr}^{-1}\mbox{Mpc}^{-3}]$, very close to the above.
The obtained $\rho_{\rm SFR}(\mbox{dust})$ is slightly larger than the local 
SFR density estimated from direct FUV radiation 
(without dust attenuation correction), 
$\rho_{\rm SFR}(\mbox{FUV}) = 8.3 \times 10^{-3} [M_\odot \,\mbox{yr}^{-1}
\mbox{Mpc}^{-3}]$ (\citealt{takeuchi05c}; see also \citealt{schiminovich05}).
Hence, 58.5~\% of the star formation is obscured by dust in the Local 
Universe.
This is in very good agreement with that of \citet{takeuchi05c}, but
since we reached this conclusion from the measured dust SED of the Local 
Universe, we could put a firmer basis on their conclusion by this work.
These results may be the first direct estimate of the dust luminosity in the
Local Universe, and should be tested by forthcoming large area survey in the
FIR by e.g., ASTRO-F.\footnote{
URL: {\tt http://www.ir.isas.ac.jp/ASTRO-F/index-e.html}.}

\section{Conclusion}\label{sec:conclusion}

We analyzed the FIRBACK $170\;\mu$m galaxy sample to obtain the 
local luminosity function (LF) of galaxies.
We constructed a flux-limited sample with $S_{170} \ge 0.195\;\mbox{Jy}$
and $z < 0.3$ from the survey, which consists of 55 galaxies.

The overall shape of the 170-$\mu$m LF is quite different from that of 
the total 60-$\mu$m LF \citep{takeuchi03b}:
the bright end of the LF declines more steeply than that of 
the 60-$\mu$m LF.
This behavior is quantitatively similar to the LF of the cool subsample
of the {\sl IRAS} PSC$z$ galaxies.
The bright end is also similar to that of the submillimeter LF of 
\citet{dunne00}.

We also estimated the strength of the evolution of the LF by assuming the 
pure luminosity evolution $L(z) \propto (1+z)^Q$.
We obtained $Q=5.0^{+2.5}_{-0.5}$ which is similar to the value obtained
by recent {\sl Spitzer} observations \citep{perez_gonzalez05,lefloch05}, 
in spite of the limited sample size.

Then, integrating over the 170-$\mu$m LF, we obtained the local luminosity 
density at $170\;\mu$m, $\rho_L(170\mu\mbox{m})\; [L_\odot \mbox{Mpc}^{-3}]$.
If we assume the above strong luminosity evolution $L(z) \propto (1+z)^5$, 
the value is $5.2 \times 10^{7} h \; [L_\odot \mbox{Mpc}^{-3}]$,
which is a considerable contribution to the local FIR luminosity density.

By summing up the other MIR/FIR data, we obtained the total dust luminosity 
density in the Local Universe.
We obtained $\rho_L(\mbox{dust})=1.1 \times 10^{8}h\;
[L_\odot \mbox{Mpc}^{-3}]$ without the evolution correction, and 
$\rho_L(\mbox{dust})=1.1 \times 10^{8}h\;
[L_\odot \mbox{Mpc}^{-3}]$ with correction.

Lastly, based on $\rho_L(\mbox{dust})$, we estimated the cosmic star 
formation rate (SFR) density hidden by dust in the Local Universe.
We took into account the dust emission heated by old
stellar population, and obtained $\rho_{\rm SFR}(\mbox{dust}) \simeq
1.1\mbox{--}1.2\, h \times 10^{-2} \; [M_\odot \,\mbox{yr}^{-1}
\mbox{Mpc}^{-3}]$.
Comparing with the SFR density estimated form FUV observation
\citep{takeuchi05c}, we found that 58.5~\% of the star formation is 
obscured by dust in the Local Universe.

It will be important to examine our local LF by a large area survey of 
the Local Universe.
The ASTRO-F project promises to provide a local large sample of FIR 
galaxies at $\simeq 50\mbox{--}170\;\mu$m.
For the evolutionary status of the FIR galaxies, the FIR data of 
{\sl Spitzer} will be very important to examine the present result.

\section*{Acknowledgments}

We are grateful for the anonymous referee for careful reading and useful 
comments which improved the clarity of this manuscript.
We also thank V\'{e}ronique Buat for fruitful discussions.
TTT and TTI have been supported by the JSPS 
(TTT: Apr.\ 2004--Dec.\ 2005; TTI: Apr.\ 2003--Mar.\ 2006).

\appendix

\section{Tables of the nonparametric luminosity function}


\begin{table*}
\centering
  \caption{The 170-$\mu$m luminosity function with $\beta=1.0$, estimated 
    with Lynden-Bell's $C^-$ method.}
\label{tab:lf_1.0}
  \begin{tabular}{lccc}
    \hline
    ${\log L_{170}}^{\mathrm{a}}$ & 
      $\phi(L_{170})$ & 
      $\phi(L_{170})^{\rm upper}$ &
      $\phi(L_{170})^{\rm lower}$ \\
    ~ & 
      $h^3\;[\mbox{Mpc}^{-3}\mbox{dex}^{-1}]$ & 
      $h^3\;[\mbox{Mpc}^{-3}\mbox{dex}^{-1}]$ & 
      $h^3\;[\mbox{Mpc}^{-3}\mbox{dex}^{-1}]$ \\
    \hline
 9.15 & $4.89 \times 10^{-2}$ & $6.20 \times 10^{-2}$ & $0.00 \times 10^{+0}$\\
 9.45 & $3.10 \times 10^{-2}$ & $4.96 \times 10^{-2}$ & $1.24 \times 10^{-2}$\\
 9.75 & $1.73 \times 10^{-2}$ & $2.63 \times 10^{-2}$ & $7.70 \times 10^{-3}$\\
10.05 & $3.81 \times 10^{-3}$ & $5.99 \times 10^{-3}$ & $1.43 \times 10^{-3}$\\
10.35 & $1.81 \times 10^{-4}$ & $3.00 \times 10^{-4}$ & $5.95 \times 10^{-5}$\\
10.65 & $2.29 \times 10^{-5}$ & $3.85 \times 10^{-5}$ & $5.54 \times 10^{-6}$\\
10.95 & $6.87 \times 10^{-6}$ & $1.17 \times 10^{-5}$ & $0.00 \times 10^{+0}$\\
    \hline
  \end{tabular}
  \begin{list}{}{}
   \item[$^{\mathrm{a}}$] Units of $L_{170}$ is $h^{-2}\;[\;L_\odot]$, and
   the bin width is $0.3\;\mbox{dex}$. Tabulated log luminosities 
   represent the bin center.
  \end{list}
\end{table*}


\begin{table*}
\centering
  \caption{Same as Table~\ref{tab:lf_1.0} but for $\beta=0.0$.}
\label{tab:lf_0.0}
  \begin{tabular}{lccc}
   \hline
    $\log L_{170}$ & 
      $\phi(L_{170})$ & 
      $\phi(L_{170})^{\rm upper}$ &
      $\phi(L_{170})^{\rm lower}$ \\
    ~ & 
      $h^3\;[\mbox{Mpc}^{-3}\mbox{dex}^{-1}]$ & 
      $h^3\;[\mbox{Mpc}^{-3}\mbox{dex}^{-1}]$ & 
      $h^3\;[\mbox{Mpc}^{-3}\mbox{dex}^{-1}]$ \\
   \hline
 9.15 & $3.33 \times 10^{-2}$ & $4.65 \times 10^{-2}$ & $0.00 \times 10^{+0}$\\
 9.45 & $5.12 \times 10^{-2}$ & $7.17 \times 10^{-2}$ & $3.10 \times 10^{-2}$\\
 9.75 & $1.69 \times 10^{-2}$ & $2.56 \times 10^{-2}$ & $7.42 \times 10^{-3}$\\
10.05 & $4.29 \times 10^{-3}$ & $6.80 \times 10^{-3}$ & $1.69 \times 10^{-3}$\\
10.35 & $1.89 \times 10^{-4}$ & $3.16 \times 10^{-4}$ & $5.95 \times 10^{-5}$\\
10.65 & $4.04 \times 10^{-5}$ & $7.00 \times 10^{-5}$ & $7.20 \times 10^{-6}$\\
10.95 & $1.20 \times 10^{-5}$ & $2.01 \times 10^{-5}$ & $0.00 \times 10^{+0}$\\
   \hline
 \end{tabular}
\end{table*}


\begin{table*}
\centering
  \caption{Same as Table~\ref{tab:lf_1.0} but for $\beta=-0.5$.}
\label{tab:lf_0.5}
  \begin{tabular}{lccc}
   \hline
    $\log L_{170}$ & 
      $\phi(L_{170})$ & 
      $\phi(L_{170})^{\rm upper}$ & 
      $\phi(L_{170})^{\rm lower}$ \\
    ~ & 
      $h^3\;[\mbox{Mpc}^{-3}\mbox{dex}^{-1}]$ & 
      $h^3\;[\mbox{Mpc}^{-3}\mbox{dex}^{-1}]$ & 
      $h^3\;[\mbox{Mpc}^{-3}\mbox{dex}^{-1}]$ \\
   \hline
 9.15 & $3.37 \times 10^{-2}$ & $4.65 \times 10^{-2}$ & $0.00 \times 10^{+0}$\\
 9.45 & $5.14 \times 10^{-2}$ & $7.21 \times 10^{-2}$ & $3.10 \times 10^{-2}$\\
 9.75 & $1.67 \times 10^{-2}$ & $2.54 \times 10^{-2}$ & $7.45 \times 10^{-3}$\\
10.05 & $4.12 \times 10^{-3}$ & $6.57 \times 10^{-3}$ & $1.53 \times 10^{-3}$\\
10.35 & $1.56 \times 10^{-4}$ & $2.63 \times 10^{-4}$ & $4.51 \times 10^{-5}$\\
10.65 & $5.45 \times 10^{-5}$ & $9.37 \times 10^{-5}$ & $1.11 \times 10^{-5}$\\
10.95 & $1.40 \times 10^{-5}$ & $2.40 \times 10^{-5}$ & $0.00 \times 10^{+0}$\\
   \hline
 \end{tabular}
\end{table*}

In this section, we present numerical tables of the nonparametric
luminosity function (LF) of the FIRBACK $170\;\mu$m galaxy sample
obtained with the $C^-$-method.
We only tabulate the LF directly estimated from the original data, and
the flux measurement errors are not included in these LFs. 

\section{Comparison between the $1/V_{\rm max}$- and $C^-$-luminosity 
functions}

\begin{figure}
\centering\includegraphics[width=0.45\textwidth]{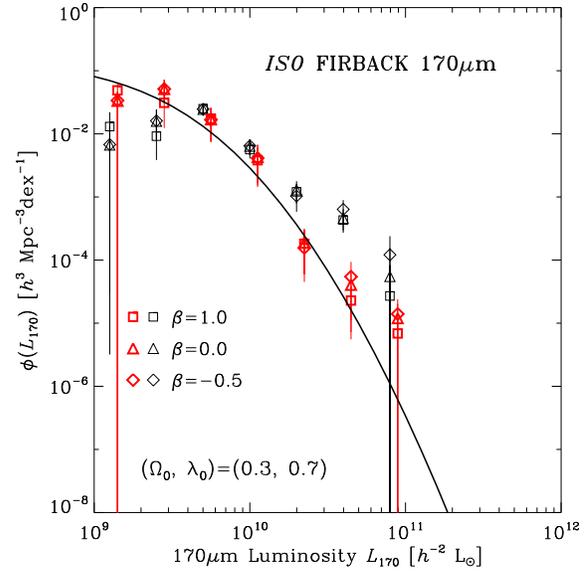}
\caption{Comparison of the luminosity functions of our {\sl ISO} 
170~$\mu$m galaxy sample by $1/V_{\rm max}$ and $C^-$ estimators.
Thick and thin symbols represent the $C^-$ nonparametric LFs, respectively.
The signification's of the symbols are the same as Figure~5
The symbols for $1/V_{\rm max}$ estimates are shifted with 0.05~dex 
for th purpose of visual clarity.
}\label{fig:lf_vmax}
\end{figure}

In this section, we examine the problem of the classical $1/V_{\rm max}$ 
estimator (Schmidt 1968; Eales 1993).
Comparison of the luminosity functions of our {\sl ISO} 
170~$\mu$m galaxy sample by $1/V_{\rm max}$ and $C^-$ estimators is shown in
Figure~\ref{fig:lf_vmax}.
In Figure~\ref{fig:lf_vmax}, the $1/V_{\rm max}$ LFs are shifted to the 
left with $0.05\;{\rm dex}$ to make
them easy to see, but the actual bin centers are exactly the same as those of
$C^-$ LFs.
It is impressive that there is a large difference between the $1/V_{\rm max}$ 
and $C^-$ LFs.
At the fainter side, these two LFs are consistent with one another within the
error bars, though the faint end of the $1/V_{\rm max}$ LFs tend to be
slightly underestimated.
In contrast, the bright end is completely different: $1/V_{\rm max}$ method
gives much flatter LFs than $C^-$ method.
We should recall that the parametric method and the $C^-$ method are both 
insensitive to density fluctuation, while $1/V_{\rm max}$ method is unbiased
only for the spatially homogeneous sample, as extensively examined by 
Takeuchi et al.~(2000).
Then, the most plausible explanation of this discrepancy may be due to the
existence of a density enhancement, corresponding to the luminosity 
$L_{170} \ga 3\times 10^{10}\;L_\odot$.

{}To understand more clearly, we recall the redshift distribution of 
the present sample (Figure~7).
The distribution of ELAIS N1 galaxies is not very far from that expected
by the LF, while that of South Marano galaxies are very different.
Reflecting the luminosity distribution of this field, which is heavily 
inclined to luminous galaxies, the distribution has no apparent peak, 
but has a widely spread shape toward redshifts up to $\sim 0.3$.
This heavy high-$z$ tail is superposed to the tail of the ELAIS N1 field
makes a significant bump at $z\simeq 0.2$.
Since the $1/V_{\rm max}$ method assumes a spatially homogeneous source
distribution, this excess is too much exaggerated through the number
density estimation, and results in the overestimation of the corresponding 
luminosity bins.

\end{document}